# Grappling with the Scale of Born-Digital Government Publications: Toward Pipelines for Processing and Searching Millions of PDFs

*Forthcoming in the International Journal of Digital Humanities*

Benjamin Charles Germain Lee
The University of Washington
Paul G. Allen School for Computer Science
& Engineering
Seattle, Washington, USA
ORCID: 0000-0002-1677-6386

Trevor Owens
American University
Public Historian in Residence
Library of Congress
Head, Digital Content Management
Washington, D.C., USA
ORCID: 0000-0001-8857-388X

**Abstract:** Official government publications are key sources for understanding the history of societies. Web publishing has fundamentally changed the scale and processes by which governments produce and disseminate information. Significantly, a range of web archiving programs have captured massive troves of government publications. For example, hundreds of millions of unique U.S. Government documents posted to the web in PDF form have been archived by libraries to date. Yet, these PDFs remain largely unutilized and understudied in part due to the challenges surrounding the development of scalable pipelines for searching and analyzing them. This paper utilizes a Library of Congress dataset of 1,000 government PDFs in order to offer initial approaches for searching and analyzing these PDFs at scale. In addition to demonstrating the utility of PDF metadata, this paper offers computationally-efficient machine learning approaches to search and discovery that utilize the PDFs' textual and visual features as well. We conclude by detailing how these methods can be operationalized at scale in order to support systems for navigating millions of PDFs.

## Declarations

**Funding:** Benjamin Charles Germain Lee's contributions are based upon work supported by the National Science Foundation Graduate Research Fellowship under Grant DGE-1762114.

**Conflicts of interest:** On behalf of all authors, the corresponding author states that there is no conflict of interest.

**Availability of data and material:** We utilize the Library of Congress's 1000 .gov PDF Dataset, a public domain dataset. The .zip download is available here, and descriptions are available here and here.

**Acknowledgments:** We thank Mark Phillips and Abbie Grotke for their input on this project, as well as the LC Web Archiving Team for their work to archive government PDFs.

## Introduction

### Scale of Government Publication Collecting

Over the past three decades, major efforts devoted to amassing digital corpora at the scale of millions and even billions of items have flourished. Such initiatives include the United States's National Digital Newspaper Program, which has digitized and made available to date over 18 million pages of historic American newspapers as part of the Chronicling America collection (National Digital Newspaper Program 2021); HathiTrust, which has digitized 17 million total volumes amounting to over 6 billion pages of content as part of their Digital Library (HathiTrust Digital Library 2021); and the Internet Archive, which has archived and made available 629 billion web pages through the Wayback Machine (Internet Archive 2021). Such initiatives have been bolstered by methodologies ranging from optical character recognition for automatedly transcribing text to crowdsourcing efforts for improving metadata.

However, the success of these initiatives has presented its own challenge: how do we navigate and analyze these collections at scale? For example, in 2006, digital humanities scholar Gregory Crane opened a special issue of D-Lib magazine with an essay entitled, "What do you do with a Million Books?" (Crane 2006). Prompted in part by questions raised by the launch of Google Books, this piece - and a substantial amount of subsequent digital humanities scholarship in the last decade and a half - have explored and developed methods that can support computational approaches to working with the wealth of digitized corpora at the scale of millions of books.

Yet, the scale of millions of digitized books represents just the tip of the iceberg for the needs of digital scholarship and digital library development. As leading digital history scholar Roy Rosenzweig had noted as early as 2003, when it comes to born-digital primary sources for scholarship, it is essential to develop methods and approaches that will scale to tens and hundreds of millions and even billions of digital resources (Rosenzweig, 2003). We have now arrived at the point where developing approaches for this scale is necessary. As an example, the Library of Congress web archives now contain more than 20 billion individual digital resources.

While the challenge of working with a million books remains daunting, basic information about each of those books existed in structured records created by librarians over the course of hundreds of years. However, with born-digital collections - including the contents of web archives - expert-created metadata is less prevalent. We therefore not only need to address the challenges of scale but also must do so in a context in which we have far less direct bibliographic control of the material at hand. As part of investigating this broad and challenging area, we have initiated an exploratory project to attempt to work with a major and significant subset of born-digital content: PDF documents archived from U.S. Government websites. In what follows, we briefly explain the scale, scope, and significance of this material and demonstrate the results of a pilot project to establish a pipeline for processing and enabling access to this kind of material that can be useful for library and archives managers, as well as end users.

## U.S. Government PDFs as a Case Study

The web has dramatically changed the nature of government publishing. Historically, units in governments like the U.S. Government Publishing Office played a key role in producing a majority of official government publications. Relatively consistent metadata for individual publications were created, and copies of publications were distributed through networks of libraries. However, as government agencies began to establish web presences in the 1990s, and as desktop software supporting the production of high quality published documents became widely available, a diverse range of government agencies and departments began to directly and frequently publish and distribute publications online. These efforts led to an explosive growth of the quantity of government publications, which in turn resulted in a breakdown of the general approach by which consistent metadata were produced for such publications and by which Libraries acquired copies of them. At the same time, these efforts caused a bifurcation between archival approaches to tracking and managing records and more library-focused approaches to publications. Substantive progress has been made in collecting and preserving this material, an overview of which can be found in the Preservation of Electronic Government Information (PEGI) Project's May 2019 National Forum Summary and Report (Halbert, et al., 2019). However, access to this material remains complex and challenging.

A key result of this progress is that approaches to web archiving, bulk crawling, and enabling replay of archived web content have become primary methods by which this content is now being preserved and made accessible. With that noted, the default method for access to web archives involves navigating to individual URLs through the re-rendered presentation of a given website. In many cases, a user must either know *a priori* exactly what they are looking for when they access archived digital resources or traverse the entirety of an archived website to find an individual digital resource of interest.

While all the files on an archived website may be of interest to users, it is worth noting that PDFs perform very specific aspects of "document-ness" that make them of key interest (Owens, 2014). Unlike HTML, PDFs are deliberately created to represent fixed layout and fixed pagination. Unlike word processing documents such as .doc files, PDFs are intended to, by default, be fixed and uneditable. These affordances make PDFs interesting as resources that relate directly to longstanding traditions in government publishing. Moreover, these affordances specific to PDFs can be leveraged as methods of enabling access and management of this content. In that context, work to extract and enhance access to embedded resources in web archives has demonstrated the value of this approach (Phillips & Murray, 2013). When PDFs were extracted from archived government websites and made discoverable as individual independent resources, they became easier for users to discover, use, and cite (Phillips, 2018). Analysis of datasets containing these files can serve as the basis for experiments with machine learning that can scale out the results of human judgments based on features of the documents (Fox & Phillips, 2019).

As a point of entry to this research, we have conducted experiments with a dataset that samples 1,000 PDFs archived from U.S. Government websites. The dataset includes a range of document types, including lengthy reports, press releases, large format maps, and structured tabular data. In exploring this data, we focus on determining what types of computationally derived metadata would be useful - potentially in combination - to support characterization, description, and search & discovery for these kinds of web-archived PDFs in bulk. Of key relevance to this project, this set of files is randomly sampled from a corpus that in 2019 contained more than 40 million unique PDFs (Thomas & Dooley, 2019). That is, working from this dataset makes it possible to extrapolate to this 40 million file corpus, and ostensibly to similar corpora of PDFs archived from the web.

While previous work has demonstrated the potential to engage with textual processing of these kinds of materials, we are specifically interested in exploring the extent to which the visual aspects of these documents could be used as a way to sort and cluster them. Of note, work with digitized newspapers has demonstrated that there is considerable potential for computer vision techniques as ways to identify not only visual information but also structured textual information and formatting like poetic verse and headlines published in historical newspapers (Lee et al. 2020; Lorang et al., 2015). Moreover, while large-scale search engines for scientific research papers such as Semantic Scholar support machine learning-aided text search over tens of millions of PDFs, these systems focus on specific features of scientific papers, rather than the heterogeneous setting of unfiltered PDFs (Ammar 2018). In what follows, we present our multimodal analysis of the 1,000 PDFs and illustrate how this could ultimately be applied to a much larger corpus of tens of millions of PDFs, which we will scale up to in future projects.

**The Dataset**

The dataset used in this study resulted from exploratory efforts by the Library of Congress's Web Archiving Team in 2018 (Library of Congress Web Archiving Program, 2019). This dataset of 1,000 PDF files was generated from indexes of the web archives, which were used to derive a random list of 1,000 items identified as PDF files and hosted on .gov domains. The set includes 1,000 unique PDF files and minimal metadata about these PDFs, including links to locations where they can be directly accessed in the Library of Congress web archive. The goal of the work was to explore the contents of the corpus of the web archives through analysis of the indexes of the harvested web content, as stored in CDX files (The CDX File Format, 2017). These indexes were used for initial analysis rather than the archived content, since that content, stored in W/ARC container files, presents significant challenges due to large size and high processing requirements (WARC, Web ARChive file format, 2009). The CDX indexes used in this initial analysis were 6 TB in size, whereas the web archive content in WARC files constituted nearly 1.5 PB at the time of analysis (Johnston, 2019). Most of the information that is provided alongside the files was extracted by the Web Archiving Team, through a process called CDX Line Extraction. For the purpose of processing the content, the only sources of materials used were the PDF files themselves. That is, there is no individual descriptive information about these files beyond the content itself and the URLs from which the content was archived.

## Modes of Metadata Generation to Support Search & Discovery

In this section, we describe our pipeline for processing the 1,000 government PDFs, as well as findings from our analysis built on top of this pipeline. The pipeline represents a conceptual model of how to integrate and interrelate 1) basic metadata like base URLs and page counts for individual PDFs with 2) text analysis metadata generated about each PDF and 3) visual analysis of the PDF files.

### The Pipeline

In Figure 1, we show a diagram of our pipeline for processing the 1,000 government PDFs. For each PDF, we begin by parsing its metadata into tabular data including features such as file size and URL. We then process the PDFs into visual and textual features as follows. First, we split the PDF into individual pages. We then convert each page to a JPEG using the command line tool ImageMagick and generate image embeddings by feeding the image into a ResNet-50 backbone and extracting the 2,048-dimensional vector from the penultimate layer; we describe this process in more detail in "Method 3: Visual Analysis of PDF Content" (The ImageMagick Development Team, 2021; He et al. 2016). We also convert each page to a text file using pdf2text and then generate TF-IDF featurizations using scikit-learn (Buitinck et al. 2011). The motivation behind and utility of these tabular, visual, and textual features are described in more detail in the following three respective sections.

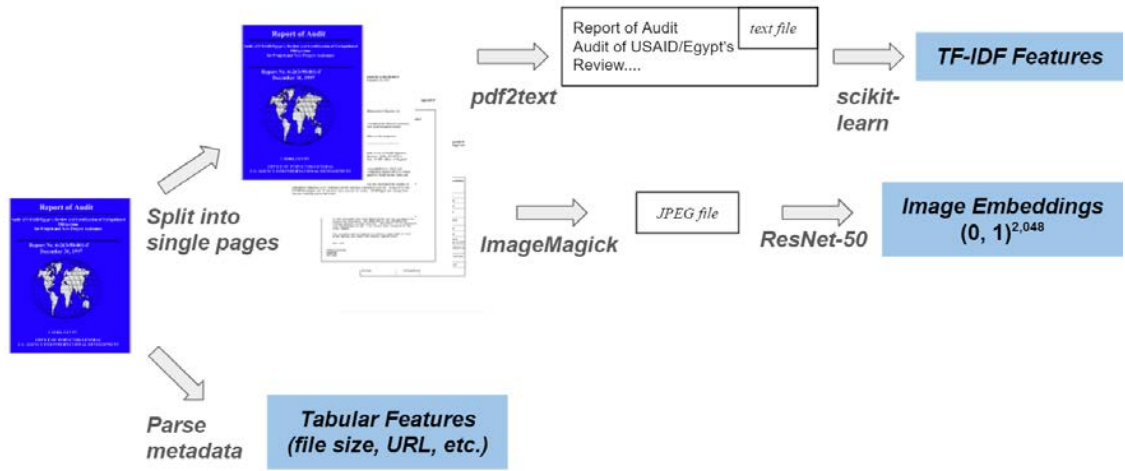

**Figure 1.** A diagram showing our preprocessing pipeline.

## Method 1: Metadata

The PDF dataset contains basic metadata produced from analysis of the files and the URLs from which they originated. In this section, we briefly review potential value of this basic metadata for search and discovery.

### Base URLs

Every PDF in the dataset was retrieved from the web. As a result, every file must have a URL from which it was made available online. In practice, the base URL for each of those files is a valuable resource for identifying what government entity made the document available. The 1,000 PDFs are associated with a total of 216 individual base URLs. 62% of the PDFs are associated with the top 20 of those base URLs presented in the figure and table below.

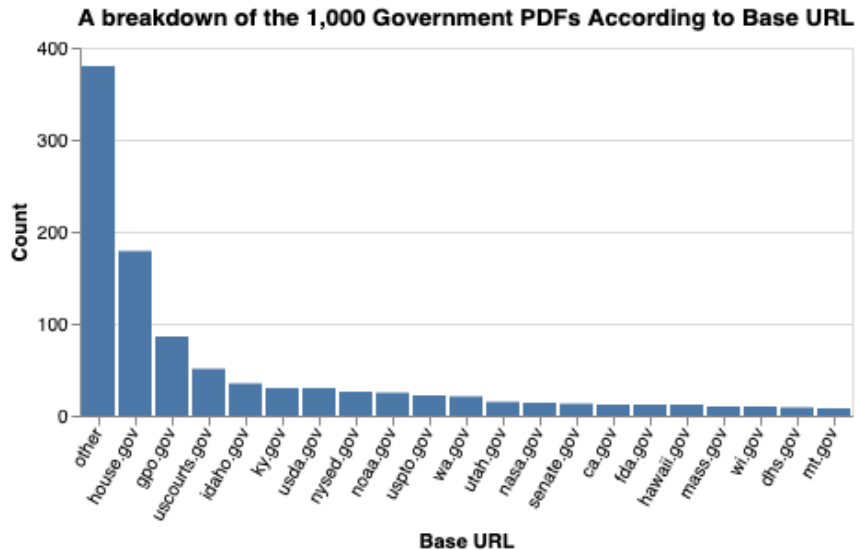

**Figure 2.** A bar graph showing the 1,000 government PDFs broken down by base URL.

| Agency | Base URL | Number |
|---|---|---|
| United States House of Representatives | house.gov | 179 |
| United States Government Publishing Office | gpo.gov | 86 |
| United States Courts | uscourts.gov | 51 |
| State of Idaho | idaho.gov | 35 |
| State of Kentucky | ky.gov | 30 |
| United States Department of Agriculture | usda.gov | 30 |
| New York State Education Department | nysed.gov | 26 |
| National Oceanic and Atmospheric Administration | noaa.gov | 25 |
| United States Patent and Trademark Office | uspto.gov | 22 |
| State of Washington | wa.gov | 21 |
| State of Utah | utah.gov | 15 |
| National Aeronautics and Space Administration | nasa.gov | 14 |
| United States Senate | senate.gov | 13 |
| State of California | ca.gov | 12 |
| U.S. Food and Drug Administration | fda.gov | 12 |
| State of Hawaii | hawaii.gov | 12 |
| Commonwealth of Massachusetts | mass.gov | 10 |
| State of Wisconsin | wi.gov | 10 |
| U.S. Department of Homeland Security | dhs.gov | 9 |
| State of Montana | mt.gov | 8 |
| Other | Other | 380 |

**Table 1**. A breakdown of the 1,000 government PDFs according to base URL.

As evident from Figure 2 and Table 1, most of the PDFs originate from the sites of large U.S. federal agencies, sites for individual U.S. state governments, and those associated with the judicial and legislative branches of the U.S. federal government. Given the consistency of this field, the straightforward ability to associate a base URL with agencies and states, and the value of information about publisher/distributor metadata, this field is invaluable in supporting search and discovery as faceted metadata for quick filtering and sorting.

**PDFs by Page Count**

In preparation of the dataset, the Library of Congress web archiving team ran the extracted files through Apache Tika to generate additional metadata. One of the fields of data they produced is the page count for each PDF file. From the 1,000 PDF sample, 987 files (98.7%) returned a page number value. The remaining 13 files are likely either corrupt in some fashion or, in some instances, password protected. As a result, nearly all of the PDF files in the dataset have page count information. Table 2 provides information on the ranges of pages associated with PDFs in the dataset.

| Number of Pages | Count of PDFs | Total pages in Corpus |
|---|---|---|
| 1 | 327 | 327 |
| 2-5 | 394 | 1,074 |
| 6-20 | 176 | 1,755 |
| 21-100 | 62 | 2,761 |
| 100+ | 28 | 6,827 |
| Null | 13 | 0 |

**Table 2**. A breakdown of the 1,000 government PDFs according to the number of pages.

The PDFs in this 1,000 file dataset - and presumably the broader corpus of more than 40 million PDFs - are mostly short or very short documents. From the whole set, 33% are 1-page documents and 39% are 2- to 5-page documents, meaning that a total of 72% of the documents are five pages or fewer. Document length drops off rapidly: 18% are 6- to 20-page documents, 6% are 20-100 pages, and 3% have 100 or more pages.

While the lengthy documents are a small portion of the set, they do still represent the majority of the pages present in the corpus. The full set of documents comprises 12,756 pages. Significantly, the very small portion of the corpus of lengthy documents - those with more than 100 pages in them - contain more than half of all the pages in the corpus. This means that work that approaches individual documents as units of analysis will highlight features of the small documents that make up the majority of the corpus while work that would approach each page in the corpus as an individual unit will focus much more attention on the smaller number of lengthy documents.

Document length is directly relevant to the format/genre of the documents. From review of subsets of these documents, it is clear that one page documents tend to contain features such as maps or technical schematics, while many of the 2-5 page PDFs are press releases and forms. Many of the 20-100 page documents are reports and other kinds of publications. The 100+ page documents include book length reports and publications, as well as entities such as bulk data printed to PDF. Given that the page count data are both highly consistent and relevant to characterizing the format and genre of the publications, they are likely to be of considerable value for supporting search and discovery.

**Method 2: Textual Analysis of PDF Content**

Most web archive interfaces rely on navigation of archived sites as their basic mode of access. Some smaller web archives have implemented full text keyword search methods, but given the scale of larger web archives, this is not widely implemented as a method of access and discovery for content. However, keyword search is itself not necessarily the ideal method for access to this material either. While keyword search is an intuitive and highly extensible method of search, it can also be limiting, as users are responsible for formulating keyword queries that retrieve relevant results.

In this section, we explore alternative text-based approaches for navigating and understanding collections of government PDFs. In particular, we utilize term frequency-inverse document frequency (TF-IDF) as a baseline method generating a feature representation of each document. TF-IDF can be used to generate a feature representation of a document as follows. For each term within a set of words and phrases, TF-IDF is a measure of the prevalence of the term within a document, relative to how prevalent the term is throughout the entire corpus, resulting in a value between 0 and 1 (a higher value indicates the term is more prevalent in the document relative to the corpus). A collection of TF-IDF values for the full set of words or phrases thus fully featurizes a document.

In our case study, we generate TF-IDF featurizations for all 1,000 PDFs in our collection using unigrams (i.e., considering single words only). It should be noted that TF-IDF is considered a baseline method of featurization among machine learning practitioners and is typically outperformed by neural text embedding models. However, TF-IDF features are easily interpretable, and this baseline method more than suffices for demonstrating the utility of text-based analysis of government PDFs for search and discovery.

**Key Terms Shared by PDFs Originating from the Same Base URLs**

We begin by analyzing terms associated with PDFs originating from different base URLs. In this context, we work to demonstrate how basic general purpose metadata facts about a PDF can be combined with computational methods to support characterization and access. For any given base domain, we can aggregate all PDFs within the 1,000 PDF dataset originating from this base domain and compute the centroid of their corresponding TF-IDF representations by averaging. We can then inspect the terms with the highest TF-IDF values in this centroid, which represents the "average" document associated with the base URL. In Table 3, we show the top 10 terms for eight base URLs from Table 1: uscourts.gov, usda.gov, nysed.gov, noaa.gov, uspto.gov, nasa.gov, senate.gov, and fda.gov. The surfaced terms are not only relevant to the agencies hosting the base URLs but are often also specific descriptors as well. These TF-IDF terms can be surfaced to curators in order to aid in the creation of finding aids. Moreover, they can help facilitate an end-user's exploratory search process by providing intuition for the PDFs within each domain grouping and helping the ideation of new keyword searches.

| Agency | Base URL | Top 10 TF-IDF Terms |
|---|---|---|
| United States Courts | uscourts.gov | court, RCJ, WGC, motion, CV, filed, case, district, united, states |
| United States Department of Agriculture | usda.gov | acres, none, farm, USDA, percent, sales, agriculture, farms, total, value |
| New York State Education Department | nysed.gov | students, school, indicator, target, disabilities, state, district, education, data, meets |
| National Oceanic and Atmospheric Administration | noaa.gov | NOAA, bay, forecast, sea, climate, species, N2O5, chart, marine, DLA |
| United States Patent and Trademark Office | uspto.gov | trademark, com, board, claims, appeal, patent, object, trial, virtual, john |
| National Aeronautics and Space Administration | nasa.gov | NASA, ingest, space, data, SABER, security, atmosphere, mission, spirit, prod |
| United States Senate | senate.gov | DTS, act, CME, treaty, senate, morgan, DOD, pool, industry, chairman |
| U.S. Food and Drug Administration | fda.gov | food, FDA, hearing, drug, lane, branch, irradiated, safety, studies, treatment |

**Table 3.** The top 10 TF-IDF terms for PDFs originating from different base URLs. Minor pruning has been applied to the lists to remove numbers, single characters, and redundant words sharing stems.

## Grouping PDFs Originating from the Same Base URLs

To investigate groupings of documents originating from the same base URLs, we utilize k-means clustering, a baseline method of unsupervised learning that partitions the PDFs into a pre-specified number of clusters using the TF-IDF featurizations. Doing so enables us to quickly determine relationships of PDFs within the broader grouping originating from the same base URL. In Table 4, we describe distinct clusters uncovered when applying k-means clusters to PDFs originating from five different base URLs: house.gov, uscourts.gov, usda.gov, nysed.gov, and fda.gov. For example, applying k-means clustering to the fda.gov PDFs with k=3 reveals distinct clusters of documents corresponding to food labeling requirements and questionnaires regarding medical waivers for hearing aids. Likewise, applying k-means clustering to the nysed.gov PDFs with k=3 reveals clusters of documents corresponding to special education in New York State, as well as assessments of schools (including "school report cards"). Unsupervised clustering thus reveals patterns among documents that cannot be inferred through existing faceted metadata alone. Surely, the PDFs originating from each base URL can be inspected manually in order to construct logical groupings within our 1,000 PDF dataset, thereby circumventing the need for clustering in this case. However, we include this proof-of-concept clustering analysis to offer provocations surrounding how one makes sense of a larger collection of PDFs; such methods are described in more detail in the penultimate section of this paper ("Toward a Unified Pipeline for Analyzing Government PDFs").

| Agency | Base URL | Descriptions of k-means Clusters |
|---|---|---|
| United States House of Representatives | house.gov (k=8) | Students & education; congressional press releases; healthcare; veterans |
| United States Courts | uscourts.gov | Bankruptcy court documents; RCJ-WGC documents from the Nevada United States District Court |
| United States Department of Agriculture | usda.gov | Tabular agricultural data; 2007 Census of Agriculture county reports |
| New York State Education Department | nysed.gov | Special education in New York State; assessments of schools ("school report cards") |
| U.S. Food and Drug Administration | fda.gov | Food labeling requirements; questionnaires regarding medical waivers for hearing aids |

**Table 4.** Descriptions of clusters within PDFs originating from the same base URLs, uncovered using k-means clustering.

**Method 3: Visual Analysis of PDF Content**

Government PDFs are just as much visual artifacts as they are textual. For example, within these PDFs, we find distinctive visual content, from maps to powerpoint slides to screenshots of webpages. Even in the case of text-heavy government documents, visual features confer meaning. The placement of text and fonts used may suggest specific document types (such as official records); black bars indicate redacted text in documents obtained through the Freedom of Information Act; government seals on letterhead indicate which agency produced the document. Significantly, these visual features and content provide information and context about documents that cannot necessarily be extracted from textual content alone.

The ability to search and analyze PDFs according to their visual features is thus a capacious approach: it not only augments existing efforts surrounding textual analysis but also reimagines what search and analysis can be for born-digital content. For example, only by considering these visual features is it possible to separate heavily redacted documents from documents with little text or assemble groupings of PDFs with maps. Providing methods for a researcher to search for distinctive visual motifs is a promising approach to unlocking collections according to specific, idiosyncratic features that are nonetheless crucial to their scholarship. And yet, visual search and analysis for born-digital content has remained understudied. Even webpage indexing relies primarily on textual content, and this flat representation restricts our view on what searching the web and born-digital content could be. In this section, we present some initial visual analysis of the 1,000 PDFs in the collection to not only demonstrate the utility of this approach but also offer provocations surrounding the operationalization of these approaches at scale.

As an initial approach, we extracted the front page of each of the 1,000 PDFs and compared the pages according to visual similarity. To accomplish this, we utilized a neural network known as ResNet-50 that had been trained to classify images in the ImageNet dataset (He et al. 2016; Deng et al. 2009). In particular, we fed each front page into ResNet-50 and extracted an image embedding from the penultimate layer of the model. This image embedding is a 2,048 dimensional vector that - when compared to that of another front page - quantifies the semantic similarity between the images. In this context, semantic similarity refers to high-level similarity, such as the presence of people in photographs or similar document structure, rather than just low level features, such as color or texture. With image embeddings produced for each of the front pages, we then utilized T-SNE to reduce the dimensionality from 2,048 dimensions to 2 dimensions, making it possible to visualize the pages' clustering (Maaten & Hinton, 2008).

In Figure 3, we present a visualization containing a global view of these 1,000 front pages. We have isolated seven distinct clusters: maps; tabular data/spreadsheets; issues of the federal register; Code of Federal Regulations; templated forms; letterhead and news releases; and heavily redacted first page documents. For each of the seven clusters, we provide a zoomed-in image, as well as a brief discussion of the visual similarity of this content and its potential significance as a means to support access to particular kinds of government documents.

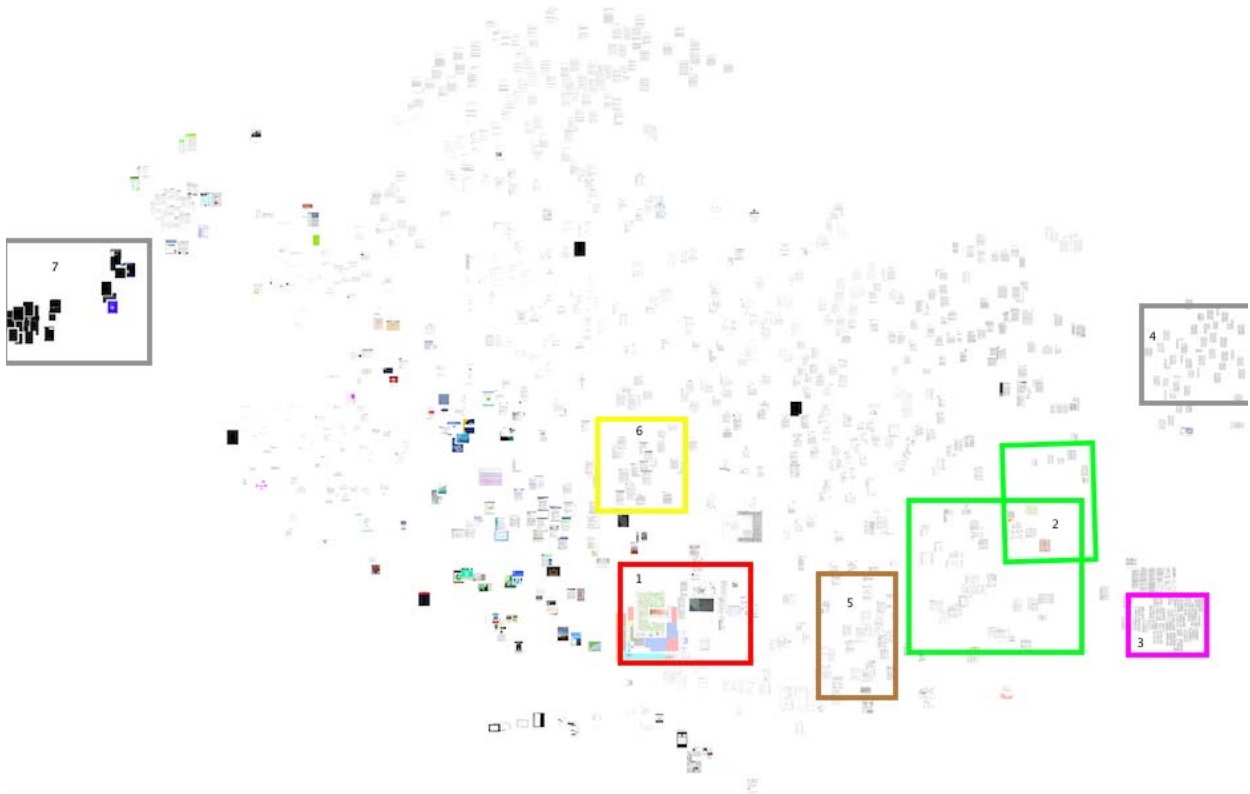

**Figure 3.** A cluster visualization of the front pages of the 1000 government PDFs in the Library of Congress dataset. The seven annotated clusters in the visualization correspond to: 1. Maps; 2. Tabular data/spreadsheets; 3. Issues of the federal register; 4. Code of Federal Regulations; 5. Templated forms; 6. Letterhead and news releases; 7. Black first page documents (i.e., heavily redacted).

## Map Content

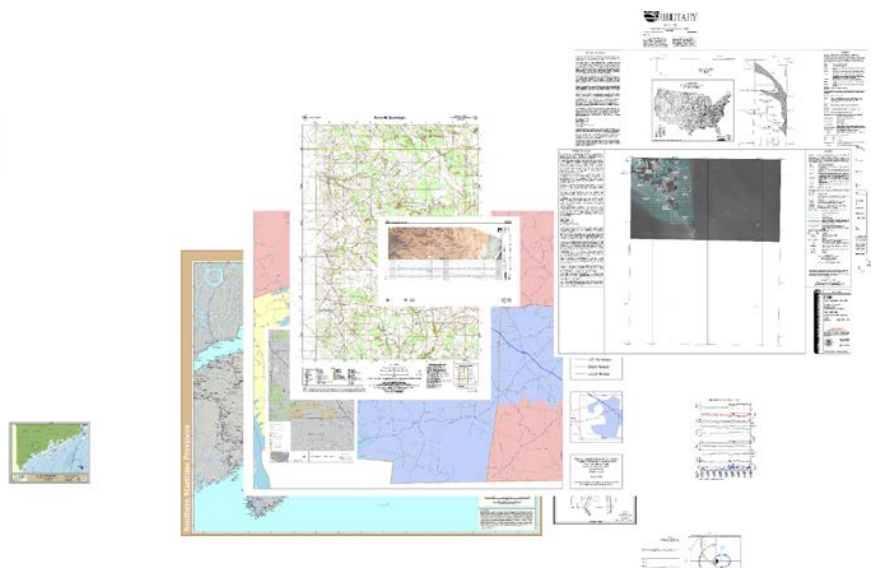

One of the most clearly visible clusters in the dataset is a series of digital map objects. Notably, official maps produced by government organizations serve a wide range of needs and purposes, and there is considerable value in being able to provide access to these geospatial resources. While there is not a large number of individual maps in this corpus, it is valuable to see that this cluster-based process is successfully identifying map similarity. Cartographic resources have often been managed separately in map libraries by experts in issues related to the geospatial aspects of these documents. To this end, it is valuable to be able to detect and classify these kinds of documents within the broader set of publications.

**Tabular Data & Spreadsheets**

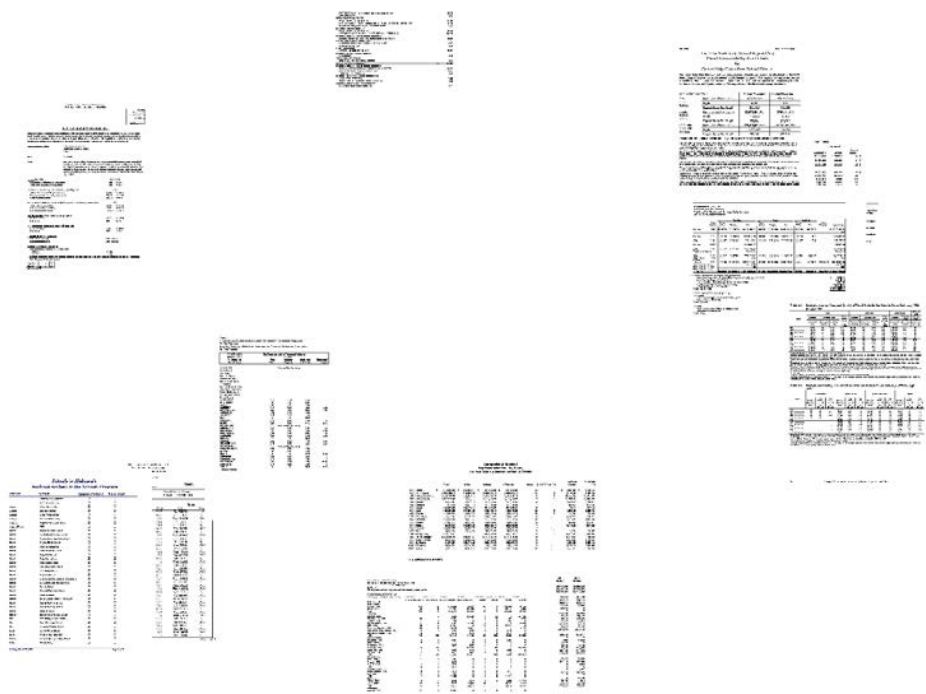

This cluster brings together documents that have highly tabular structures to them. Many of these appear to be printed to file versions of spreadsheets and similar documents. Many government publications over time have been focused on presenting this kind of tabular data and statistical information. To this end, these kinds of documents are more like datasets published by government entities than textual documents intended to be read as written works. In this context, identification of these kinds of tabular data PDFs is likely of keen interest and relevance to ongoing work with open government datasets.

**Issues of the Federal Register and the Code of Federal Regulations**

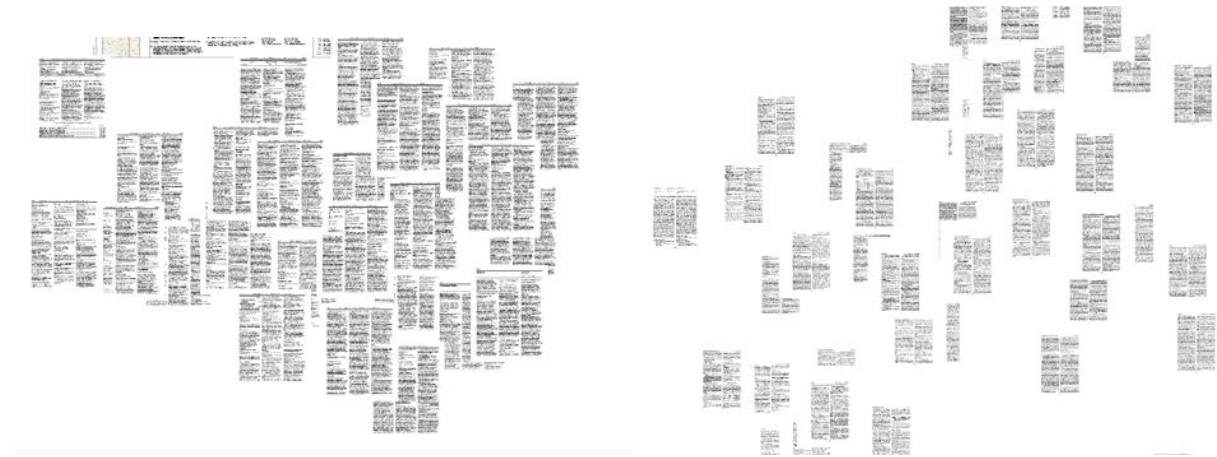

Some forms of government publishing take on very consistent templated layouts and structures and were prevalent enough in the dataset to generate their own visual clusters. In this case, the three-column layout of issues of the Federal Register and the two-column layout of the Code of Federal Regulations both come through as distinct document clusters.

**Templated Forms**

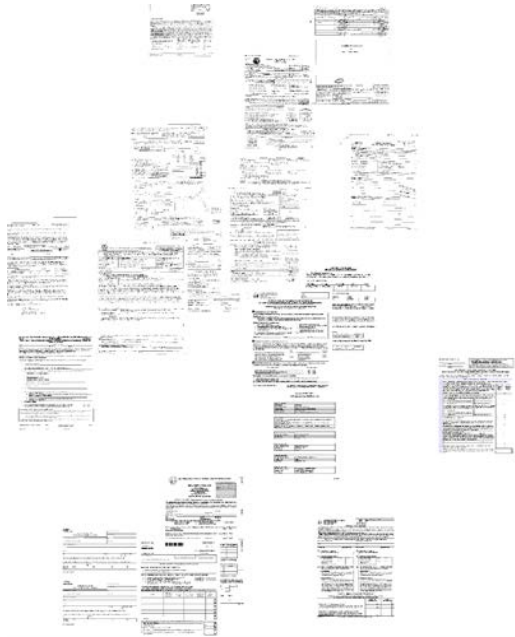

A key aspect of the function of PDFs is that they function as highly structured official government forms. In this cluster we can see a range of somewhat related templated form documents. Given how highly structured these kinds of forms are, and the extent to which all instances of a form like this will contain those same elements, we anticipate that this kind of visual clustering method will be particularly useful for templated forms in larger corpora of documents. For example, this kind of method should ultimately not only be able to detect clusters of forms but also be able to produce sub clusters of specific forms of broad interest, such as the 990 IRS tax document required for nonprofit organizations.

**News releases on letterhead**

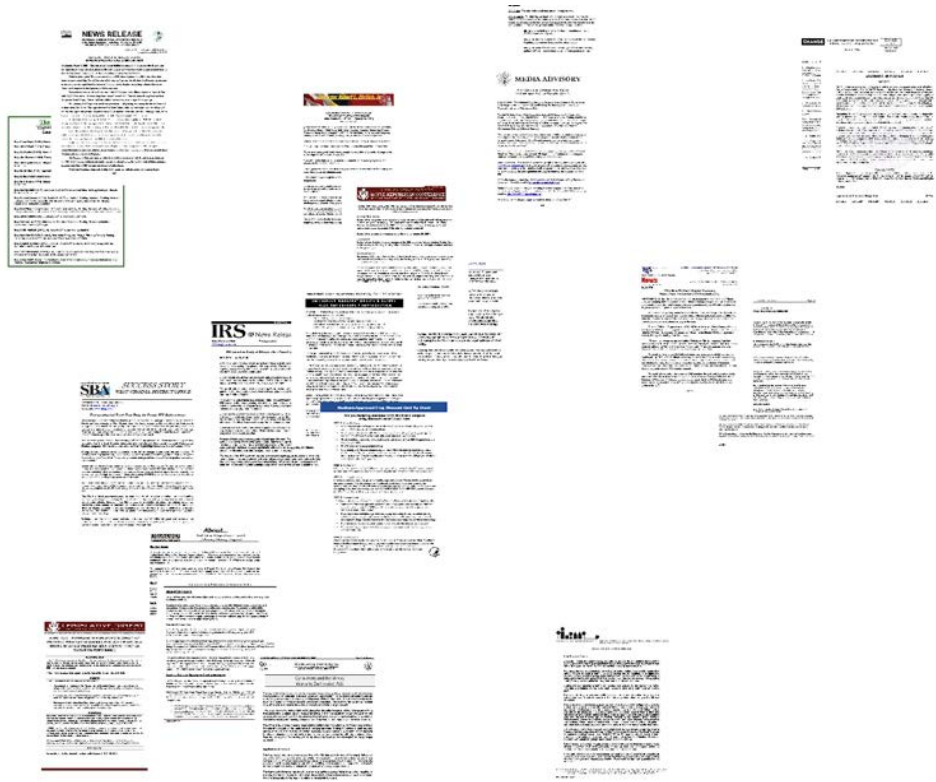

This cluster includes a varied range of documents on official agencies' letterheads published as press or news releases. Significantly, this cluster emerges to include not just documents from one individual agency but also releases from a range of different departments and agencies including the IRS and the Small Business Administration. As documents published with the main purpose of shaping media and news coverage of the work of the government, this type of cluster is of particular value to explorations of the history of public affairs and media relations in government.

**Redacted Front Pages**

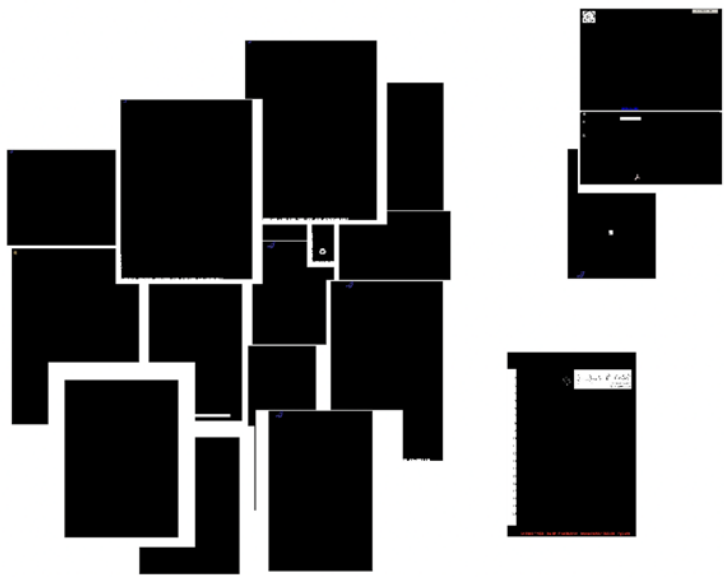

The last cluster presents itself on the display as a significant outlier but may be of interest to researchers. This set of documents has had the first pages heavily redacted and blacked out. For scholars interested in disclosures of classified or otherwise sensitive information, it may well be of particular interest to isolate exactly this type of cluster of material for analysis and research.

We conclude this section by reiterating that these methods offer substantial utility to a range of practitioners. For example, curators can utilize these approaches to assess a collection at scale, identify and label subcollections, find overabundances, and assess what is missing. Moreover, end-users such as scholars can use these methods to explore and analyze these collections.

**Toward a Unified Pipeline for Analyzing Government PDFs**

While digitized collections have already inspired the utilization of machine learning-powered methods of search and analysis in response to questions of scale, government PDFs - and web archives more generally - necessitate an even greater reliance on these methods. The 2008 End of Term Presidential Web Archive alone contains approximately ten million PDFs, and the total number of government PDFs held by the Library of Congress and the Internet Archive is in the hundreds of millions (Phillips 2008). Accordingly, in this section, we describe potential baseline paths forward for machine learning-aided search and discovery systems that empower people to search across these collections at scale. Because basic metadata facets and keyword search are standard affordances for collections of this scale, we focus on the machine learning-aided similarity search affordances presented in the previous section of this paper.

Visual search can already be operationalized across tens of millions of PDFs in a tractable manner. Image embeddings, such as the ones described in the previous section, can be pre-computed when PDFs are initially indexed and stored. These image embeddings can be utilized not only to produce cluster-based affordances but also to define additional facets in a semi-automated fashion. For example, a curator could incorporate a facet corresponding to a visual feature, such as "partially redacted document," before the deployment of a search system by utilizing a machine learner to retrieve PDFs relevant to the specific facet of interest. Moreover, facets can be generated by users on the fly. In particular, a user can train an interactive machine learner on just a handful of labeled examples in order to retrieve PDFs according to visual features of interest during the exploratory search process. For instance, one of the authors of this paper (BCGL) has provided end-users with a version of this affordance for searching 1.5 million historic newspaper photographs. As shown in Figure 4, a user can easily train a machine learner to retrieve examples of sailboats across the entire corpus of photographs by labeling a handful of examples of sailboats (Lee and Weld, 2020). The machine learner can train and predict across all 1.5 million photos in just a second, making it possible for the user to iterate quickly and generate relevant sets of images with ease. This approach can be extended to support the iterative training of machine learners for retrieving PDFs with distinct visual features across an entire corpus, such as the 2008 End of Term Presidential Web Archive, without introducing inordinate latency.[1] With this approach, a user can quickly retrieve groupings such as the ones identified in Figure 3.

Similar approaches can be utilized to facilitate text-based similarity search and recommendation. Featurizations can be pre-computed by converting PDFs to text documents upon indexing and generating TF-IDF vectors or text embeddings. Like image embeddings, these textual features are lightweight representations that can be utilized for a host of downstream tasks at scale. For example, a system could easily support similarity search and recommendation by providing interactive machine learning affordances akin to the visual similarity search affordances described in the previous paragraph. Additionally, efficient unsupervised clustering methods can be utilized to reveal new groupings of PDFs that are not captured within existing metadata. Lastly, a more expressive facet taxonomy could be incorporated into a search system utilizing scalable natural language processing methods for automated or semi-automated facet taxonomy construction (Stoica et al. 2007). When combined with keyword search and basic metadata filtering, such methods of facet construction have the capacity to drastically enrich the navigation experience.

In summary, the pre-processing pipeline presented in this paper represents a scalable solution for handling millions of PDFs. The methods articulated in this section are tractable approaches to searching and analyzing PDFs at scale, and we will report on our efforts to build out these systems and affordances in our follow-up papers.

---

[1] In particular, issues surrounding training latency can be circumvented by utilizing linear models for the interactive machine learners. Moreover, challenges surrounding memory (i.e., predicting over tens of millions of pages, each represented by a 2,048-dimensional vector) can be solved via parallelization across machines and chunking.

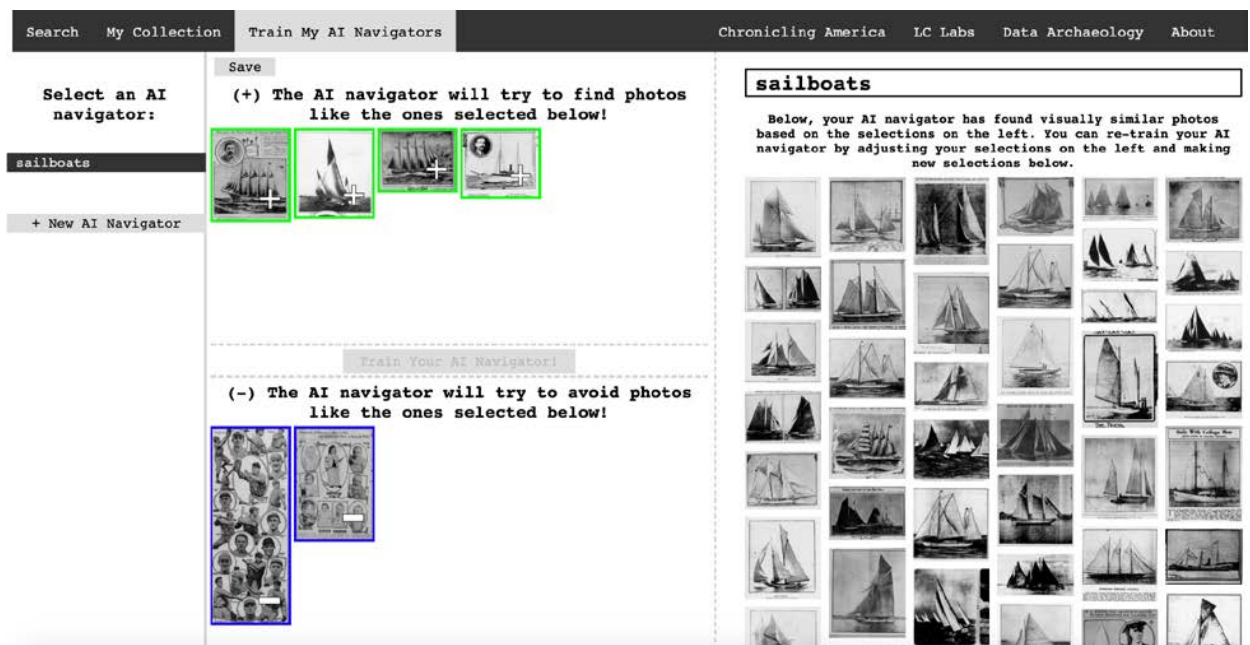

**Figure 4.** An example of an interactive machine learning interface for retrieving photographs of sailboats in the *Newspaper Navigator* search application (Lee and Weld, 2020).

## Conclusion and Future Work

Government PDFs - and web archives writ large - represent invaluable sources for scholars. Yet, affordances for searching large collections at scale remain limited and underexplored to date. In this paper, we have reported on a scalable pipeline for processing and analyzing government PDFs. Beginning with a dataset of 1,000 PDFs, we have demonstrated the utility of search methods across different modalities, including tabular metadata, textual features, and visual features. We have shown how these methods have the capacity to enrich the search experience for a range of users, from curators to scholars. We have also described how our pipeline is scalable and can be operationalized across millions of PDFs.

Accordingly, future work includes building out our pipeline, as well as visual and textual search affordances, for a larger dataset, such as the 2008 End of Term Presidential Web Archive Dataset. As our next immediate step, we plan to shift to work toward a combined corpus of tens of thousands of similar PDFs to further test these ideas and then ultimately scale up to working with millions and then tens of millions of PDFs. We intend to do so with the goal of not only demonstrating such a system in practice but also evaluating our system with real users of these systems: librarians, archivists, historians, lawyers, and other scholars. We will work with these user groups to understand how these affordances could be best utilized and implemented. Moreover, we plan to refine and improve our search affordances to enable more granular search queries, for example, retrieving all PDFs with specific features on their pages, such as the congressional seal. We also plan to explore multimodal approaches of search and analysis by learning from combined visual and textual representations. Lastly, we intend to experiment with pipelines and systems that unify different collections of PDFs in order to demonstrate generalizability and also draw insights across collections. We offer this work in pursuit of deployed machine learning-aided search and discovery systems for millions of government PDFs.